\documentclass[graybox]{svmult}

\usepackage{cite}
\usepackage{mathptmx}       % selects Times Roman as basic font
\usepackage{helvet}         % selects Helvetica as sans-serif font
\usepackage{courier}        % selects Courier as typewriter font
\usepackage{type1cm}        % activate if the above 3 fonts are
\usepackage{makeidx}         % allows index generation
\usepackage{graphicx}        % standard LaTeX graphics tool
\usepackage{multicol}        % used for the two-column index
\usepackage[bottom]{footmisc}% places footnotes at page bottom

\makeindex             % used for the subject index
\begin{document}

\title*{Quantum black holes and effective quantum gravity approaches}
\author{Xavier Calmet}
\institute{ \at  Physics $\&$ Astronomy, 
University of Sussex,   Falmer, Brighton, BN1 9QH, UK  \email{x.calmet@sussex.ac.uk}}
\maketitle

\abstract*{blabla
}

%\section{Low scale quantum gravity and black holes at colliders}
One of the most exciting developments in theoretical physics in the last 20 years has been the realization that the scale of quantum gravity could be in the TeV region instead of the usually assumed $10^{19}$ GeV. Indeed, the strength of gravity can be affected by the size of potential extra-dimensions  \cite{ArkaniHamed:1998rs,Antoniadis:1998ig,Randall:1999ee,Gogberashvili:1998vx} or the quantum fluctuations of a large hidden sector of particles \cite{Calmet:2008tn}. A dramatic signal of quantum gravity in the TeV region would be the production of small black holes in high energy collisions of particles at colliders.  The possibility of creating small black holes at colliders has led to some wonderful theoretical works on the formation of black holes in the collisions of particles.

Long before studying the production of such black holes in the high energy collisions of particles became fashionable, in the 1970's Penrose proved that a closed trapped surface forms when two shockwaves traveling at energies much larger than the Planck scale even when the impact parameter is non-zero. Unfortunately, he never published his work. The result was independently rediscovered by Eardley and Giddings in 2002 \cite{Eardley:2002re} when the high energy community started to discuss the formation of black holes at colliders. Earlier estimate of the production cross section had been done using the hoop conjecture. Some did not trust the hoop conjecture, thinking that in the collision of particles the situation was too asymmetrical to trust this conjecture. The paper of Eardley and Giddings settled the issue. Proving the formation of a closed trapped surface is enough to establish gravitational collapse and hence the formation of a black hole. This work was extended by Hsu \cite{Hsu:2002bd} into the semi-classical region using path integral methods. One could thus claim with confidence that black holes with masses 5 to 20 times the Planck scale, depending on the model of quantum gravity, could form in the collision of particles at the CERN LHC is the Planck scale was low enough. Early phenomenological studies can be found in \cite{Dimopoulos:2001hw,Giddings:2001bu,Feng:2001ib,Anchordoqui:2003ug,Anchordoqui:2001cg,Anchordoqui:2003jr,Hossenfelder:2001dn}.

However, it is obvious that even if the Planck scale was precisely at 1 TeV not many semi-classical black holes could be produced at the LHC since the center of mass energy of the collisions between the protons was at most of 8 TeV so far \cite{Meade:2007sz}. Even with the 14 TeV LHC, not many if any semi-classical black holes will be produced.

We thus focussed on quantum black holes, which are black holes with masses of the order of the Planck mass which could be produced copiously at the LHC or in cosmic ray experiments \cite{Calmet:2008dg,Calmet:2010vp,Calmet:2011ta,Calmet:2012cn,Calmet:2012fv,Alberghi:2013hca,Calmet:2008rv,Calmet:2012mf,Arsene:2013nca,Calmet:2010nt}. The current bound derived using LHC data on the fist quantum black hole mass if of the order of 5.3 TeV \cite{Aad:2013gma,Savina:2013eja}. Note that this bound is slightly model dependent. However, this is a clear sign that there are no quantum gravitational effects at 1 TeV. 

At the time we are writing up this paper, there is actually no sign of any physics beyond the standard model in the TeV region. It  thus seems that the hierarchy problem was a red herring; a light Higgs boson has been found, but there is no sign of new physics to stabilize the Higgs boson's mass. This is the second nail in the coffin for fine-tuning problems after the discovery of a small and non-zero cosmological constant without new physics to stabilize it.

%\section{An effective theory for quantum gravity}

Instead of trying to probe the Planck scale directly by producing small black holes directly at colliders, it is useful to think of alternative ways to probe the scale of quantum gravity. Effective field theory techniques are very powerful when we know the symmetries of the low energy action which is the case for the standard model of particle physics coupled to general relativity. Integrating out all quantum gravitational effects, we are left with an effective action which we can use to probe the scale of quantum gravity at low energies. We thus consider:
\begin{eqnarray}\label{action1} \nonumber
S \!=\! \int d^4x \, \sqrt{-g} \left[ \left( \frac{1}{2}  M^2 + \xi H^\dagger H \right)  R- \Lambda_C^4  + c_1 R^2  
 + c_2 R_{\mu\nu} R^{\mu\nu}  + L_{SM} + O(M_\star^{-2})   \right]  \\
\end{eqnarray}
 The Higgs boson $H$ has a non-zero vacuum expectation value, $v=246$ GeV and thus contribute to the value of the Planck scale:  
\begin{eqnarray}
\label{effPlanck}(M^2+\xi v^2)=M_P^2 \, .
\end{eqnarray}
The parameter $\xi$ is  the non-minimal coupling between the Higgs boson and space-time curvature. The three parameters $c_1$, $c_2$ and $\xi$ are dimensionless free parameters. The Planck scale $M_P$ is equal to $2.4335 \times 10^{18}$ GeV and the cosmological constant $\Lambda_C$ is of order of $10^{-3}$ eV.  The scale of the expansion $M_\star$ is often identified with $M_P$ but there is no necessity for that and experiments are very useful to set limits on higher dimensional operators suppressed by $M_\star$.
Submillimeter  pendulum tests of Newton's law \cite{Hoyle:2004cw} are used to set limits on $c_1$ and $c_2$. In the absence of accidental cancellations between the coefficients of the terms $R^2$ and $R_{\mu\nu} R^{\mu\nu}$, these coefficients are  constrained to be less than $10^{61}$ \cite{Calmet:2008tn}. It has been shown that astrophysical observations are unlikely to improve these bounds \cite{Calmet:2013hfa}. The LHC data can be used to set a limit on the value of the Higgs boson non-minimal coupling to space-time curvature: one finds that $|\xi| >2.6 \times 10^{15}$ is excluded at the $95 \%$ C.L.  \cite{Atkins:2012yn}. Very little is known about higher dimensional operators. The Kretschmann scalar $K = R^{\mu\nu\rho\sigma} R_{\mu\nu\rho\sigma}$ which can be coupled to the Higgs field via  $K H^\dagger H$ has been studied in \cite{Onofrio:2012zz}, but it seems that any observable effect requires an anomalously large Wilson coefficient for this operator. Clearly one will have to be very creative to find a way to measure the parameters of this effective action. This is important as these terms are in principle calculable in a theory of quantum gravity and this might be the only possibility to ever probe quantum gravity indirectly. 

%\section{What is the limitation of this effective field theory}

The standard model is very, maybe even, too successful. At what energy scale can we expect it to break down? In other words, up to what energy scale can one trust the effective theory described above? We know that this effective theory does not describe dark matter, but this could be a hidden sector of particles or maybe even primordial black holes with masses of the order of the Planck mass which would not affect the effective action and our previous conclusions. It has been recently pointed out that if gravity is asymptotically safe, the effective theory (\ref{action1}) could offer a description of nature up to arbitrarily energy scale and predict the Higgs boson's mass correctly, i.e. at 126 GeV \cite{Shaposhnikov:2009pv}. Within this framework, it is natural that instead of considering the Higgs boson as a source of the hierarchy problem, one should look at it as a  solution to another type of fine-tuning issue, namely that of the initial conditions of our universe.
The fine-tuning problematic at the beginning of our universe is very different from the fine-tuning problem
in the standard model. The fine-tuning issue in cosmology is really an initial condition problem.
Why did our universe start from such very specific initial conditions?
It has been shown in Ref.~\cite{Bezrukov:2007ep,Barvinsky:2008ia,DeSimone:2008ei,Bezrukov:2013fka} that the Higgs boson with a  non-minimal coupling to the Ricci scalar could play the role of the inflaton and thus address this problem.

However, getting the right number of e-folding requires a fairly large non-minimal coupling of the order of $10^4$.
This large non-minimal coupling is the source of a potential issue with perturbative unitarity
(see, e.g.~\cite{Lerner:2009na,Burgess:2010zq,Atkins:2010yg,Hertzberg:2010dc} and references therein).
Naively, unitarity seems to be violated at an energy scale of $M_P/\xi$ in
today's Higgs vacuum, while it would be violated at a scale  $M_P/\sqrt{\xi}$ in the inflationary
background. The breakdown of perturbative unitarity is a sign of strong dynamics or new physics which kicks
in at the scale of the breakdown of perturbative unitarity, thereby restoring unitarity.
However, both new physics and strong dynamics could jeopardize the flatness of the scalar potential
which is needed to obtain the correct number of e-folding required to explain the flatness of our universe. It was shown in 
\cite{Calmet:2013hia} that at least at one-loop the cutting relation is fulfilled which implies that perturbative unitarity is fixed by one-loop corrections. This is an example of the self-healing mechanism discussed in \cite{Aydemir:2012nz}. The implication of this calculation is that the standard model could be valid at least up to the Planck scale,  and describe particle physics and inflation in one consistent framework.

Unless quantum gravity is asymptotically free, proving or disproving this remains a calculational challenge as it is a purely non-perturbative problem, the effective theory (\ref{action1}) will certainly breakdown at the scale at which quantum gravitational effects become large. The lack of success in finding a consistent theory of quantum gravity may be an indication that gravity does not need to be quantized in the usual sense, or that we are trying to quantize the wrong degrees of freedom. The metric may be something purely classical and emergent.  Physics seems to be in a crisis again as in 1900 when Lord Kelvin said ``There is nothing new to be discovered in physics now. All that remains is more and more precise measurement''. We should just hope that, as at the start of the 20th century, we will experience a new scientific revolution. My point of view is that we may have reached the limit of what can be done within our current theoretical framework. After all, quantum field theory is still based on very classical concepts namely that of point mechanics: we specify the energy of a particle which we split into kinetic and potential energies.  The couplings and masses of the standard model are nothing but proportionality constants between the kinetic terms and the potentials for the corresponding particles. Yes, we quantize the classical theory to obtain a quantum field theory, but the underlying ideas and principles are desperately classical. This may be the reason why we have been unable to make progress and to calculate some of the fundamental constants of nature such as the coefficients of our effective action (\ref{action1}). Any progress will require some bright idea. We can hope that black holes will give us some clues of how to proceed beyond the current paradigm.

{\it Acknowledgments:}
This work is supported in part by the European Cooperation in Science and Technology (COST) action MP0905 ``Black Holes in a Violent  Universe" and by the Science and Technology Facilities Council (grant number  ST/J000477/1).

\end{document}